\newcommand{\rem}[1]{}
\documentclass{amsart}
\usepackage{amsfonts,amssymb,amsmath,amsthm,mathrsfs}
\usepackage{comment}
\usepackage{url}
\usepackage[dvips]{epsfig}
\newtheorem{thrm}{Theorem}[section]
\newtheorem{lem}[thrm]{Lemma}
\newtheorem{prop}[thrm]{Proposition}

\newtheorem{remark}[thrm]{Remark}
\theoremstyle{definition}
\newtheorem{definition}[thrm]{Definition}

\begin{document}
\author[C.~A.~Mantica and L.~G.~Molinari]
{Carlo~Alberto~Mantica and Luca~Guido~Molinari}
\address{C.~A.~Mantica: I.I.S. Lagrange, Via L. Modignani 65, 
20161, Milano, Italy -- L.~G.~Molinari (corresponding author): Physics Department,
Universit\`a degli Studi di Milano and I.N.F.N. sez. Milano,
Via Celoria 16, 20133 Milano, Italy.}
\email{carloalberto.mantica@libero.it, luca.molinari@unimi.it}
\subjclass[2010]{Primary 53B30, 53B50, Secondary 53C80, 83C15}
\keywords{Generalized Robertson-Walker spacetime, perfect fluid spacetime, 
Killing vector, conformal Killing vector, torse-forming vector, concircular vector, 
conformal Killing tensor, Robertson-Walker spacetime.}
\title[GRW spacetimes - a survey]
{Generalized Robertson-Walker spacetimes\\ - a survey -}
\begin{abstract} 
Generalized Robertson-Walker spacetimes extend the notion of Robertson-Wal\-ker spacetimes,
by allowing for spatial non-homogeneity. A survey is presented, with main focus on
Chen's characterization in terms of a timelike concircular vector. Together with their most important properties, some new results are presented.
\end{abstract}
\date{07 dec 2016}
\maketitle
\section*{Introduction}
Spacetime is the stage of present modelling of the physical world: a torsionless, time-oriented  
Lorentzian manifold $(M,g)$. % of dimension $n$.
At any point, by a suitable choice of coordinates,
the metric tensor is amenable to the diagonal form\footnote{Latin indices as $abij$ take values $0,1,...,n-1$, Greek indices as $\alpha\beta\mu\nu$
take values $1,...,n-1$.} $g_{00}=-1$, $g_{\mu\mu} =+1$, 
with vanishing first derivatives (Christoffel symbols). 
Such coordinates define a local inertial frame. Gravity is described by second derivatives, that build the Riemann curvature tensor $R_{jklm}$,  
the Ricci tensor $R_{ij}=R_{imj}{}^m$ and the curvature scalar $R=R_m{}^m$. 
Einstein's field equations are second-order in the metric tensor, with source the stress-energy tensor of ``matter" fields \cite{[Bee96],Wald}. In
geometrized units:
\begin{align}
R_{ij} - \tfrac{1}{2} R g_{ij} = 8\pi T_{ij} \label{eq_Einstein}
\end{align} 
They have been widely
tested in the weak field expansion and recently in the 
strong coupling regime, by numerically solving for the merging of two black-holes and providing the detected spectrum of gravitational waves \cite{Abbott16}.

In cosmology, the observation that space is isotropic and homogeneous on the large scale, 
selects the Robertson-Walker (RW) metrics. In $n$ dimensions it means that there are coordinates $(t,\vec x)$ such that  
$M=-I\times f^2 M^*$, where $I$ is an open interval of the real line and 
$(M^*, g^*)$ is a constant curvature Riemannian space of dimension $n-1$, 
parametrized by $\vec x$. The Riemann tensor of such submanifolds is 
\begin{align}\label{eq_constcurv}
R^*_{\mu\nu\rho\sigma} = \frac{R^*}{(n-1)(n-2)} (g^*_{\nu\sigma} g^*_{\mu\rho}-g^*_{\mu\sigma}g^*_{\nu\rho}),
\end{align}
with constant value $R^*$.\\
To study perturbations of RW metrics, homogeneity must be relaxed. A natural and wide extension was proposed in 1995
by Al\'{\i}as, Romero and S\'anchez  \cite{AliRomSan95_a,AliRomSan95_b}: 
\begin{definition}\label{def_GRW}
A Lorentzian manifold $(M,g)$ of dimension $n\ge 3$ is a Generalized Robertson-Walker (GRW) spacetime if the metric may take the form 
\begin{align}\label{eq_1.1}
 g_{ij} dx^i dx^j = -(dt)^2 + f(t)^2 g_{\mu\nu}^*(\vec x) dx^\mu dx^\nu 
\end{align}
where $t$ is time and $g^*_{\mu\nu}(\vec x)$ is the metric tensor of a Riemannian submanifold. 
\end{definition}
\noindent
A GRW spacetime is thus the warped product $-I\times f^2 M^*$, where $I$ is an open 
interval of the real line, and $(M^*,g^*)$ (the fiber) is a Rie\-man\-nian manifold of dimension $n-1$; 
$f>0$ is a smooth warping function (or scale factor).\\
1) if $M^*$ is the unit sphere $\mathbb S^{n-1}$ with its standard metric, $I=\mathbb R$, $f(t)=\cosh t$, then $M$ is the de Sitter space $\mathbb S^n_1$, of constant sectional curvature 1; \\
2) if $M^*$ is a constant curvature manifold \eqref{eq_constcurv}, then $M$ is a RW spacetime\footnote{Some authors name GRW a RW spacetime of dimension $n>4$.}.\\
A further natural generalization brings from GRW spacetimes to ``twisted spacetimes'', where the scale function $f$ in the metric
\eqref{eq_1.1} may also depend on space coordinates. Twisted spaces were introduced by Chen in 1979 \cite{Chen79}.

The existence of a globally 
defined timelike coordinate vector field $\partial_t$ makes GRW spacetimes (by definition) 
time-orientable. This endows them with a causal structure appropriate for the 
time-evolution of physical fields, propagating initial conditions given on a spacelike hypersurface. \\
The definition \eqref{eq_1.1} highlights the existence of a continuous family of spacelike hypersurfaces (foliation): for each $t\in I$ there is a ``slice" $\{t \}\times M^*$
(leaf of the foliation). A slice is a totally umbilical hypersurface of constant mean curvature: at every point of it, the average of 
the principal curvatures has the same value $H(t)=f^\prime /f$, where $t$ labels the slice.
Ref. \cite{AliRomSan95_a}, where GRW spacetimes were first introduced, opens with the following question: ``when is a complete spacelike hypersurface of constant mean curvature (in a GRW spacetime) totally umbilical and a slice?'' A large portion of the literature on GRW manifolds is focused on this and related problems, like 
curvature properties of spacelike hypersurfaces.

A different characterization of GRW spacetimes was found by Chen, in 2014. He showed that the existence of a warping
frame \eqref{eq_1.1} is equivalent to the existence of a concircular timelike vector field \cite{Chen14}. The 
theorem offers a covariant description of GRW spacetimes, alternative to the geometric description in a privileged frame. Recently, it allowed for new results on the structure of the Ricci and the Weyl tensors to be obtained, and new characterizations of 
GRW spacetimes.\\
This review aims at presenting GRW spacetimes from both viewpoints, 
which offer different advantages, with an inclination towards Chen's approach. Proofs are omitted, if available in published papers. 
This is the summary of contents:

1. Spacelike hypersurfaces; 

2. Killing vectors;

3. Chen's theorem; 

4. GRW spacetimes and conformal Killing tensors; 

5. Perfect fluid GRW spacetimes; 

6. Further characterizations of GRW spacetimes; 

7. Robertson-Walker spacetimes;

8. Other curvature conditions;

9. Remarks on imperfect fluid GRW spacetimes.\\

\noindent
Given a vector $v$, we denote $v^2=v^kv_k$; $v$ is respectively timelike, null, or spacelike if $v^2$ is negative, zero or positive. The dimension of the manifold $M$, if not specified, is $n\ge 3$.
%
%%%%%%%%%%%%%%%%%%%%%%%%%%%%%%%%%%%%%%%%% 
%
\section{\bf Spacelike hypersurfaces}
We present a selection of results on spacelike hypersurfaces
of GRW spacetimes. They are based on a variety of tools, as convexity 
properties of the warping function, inequalities for the Ricci tensor, integral identities involving curvatures $H$, $H_2$,
warping function, hyperbolic angle. 

With the metric tensor \eqref{eq_1.1}, the nonzero Christoffel symbols and the components of the Ricci tensor \cite{Sanchez99} are:
\begin{gather}
\Gamma^0_{\mu\nu} =  ff' g^*_{\mu\nu} \, , \quad \Gamma^\mu_{0\nu}  
=(f'/f) \delta^\mu_\nu \, , 
\quad \Gamma^\mu_{\nu\sigma} =\Gamma^{*\mu}_{\nu\sigma} \label{eq_Christoffel}.\\
 R_{00} = -(n-1)(f^{\prime\prime}/f), \qquad R_{0\mu}=R_{\mu 0}=0,\label{eq_3.1} \\
 R_{\mu\nu} =R^*_{\mu\nu} + g^*_{\mu\nu} [(n-2) f^{\prime 2}+  f f^{\prime\prime} ],
\end{gather}
where the suffix $*$ denotes quantities of the submanifold $(M^*,g^*)$.
The eigenvalue equation for $R^i{}_j$ has a single 
timelike eigenvector $X=(X^0,0)$, with eigenvalue 
\begin{align}\label{eq_xi1}
\xi = (n-1)\, f^{\prime\prime}/f
\end{align}
that only depends on $t$, and $n-1$ spacelike eigenvectors (with time-component equal to zero).
The eigenvalue \eqref{eq_xi1} is a frame-independent feature of the GRW spacetime.
We shall show that the choice $X^0=f$, i.e. $X=f\partial_t$, corresponds to the timelike concircular Chen's vector.

A spacelike hypersurface $\Sigma $ in $M$ is a smooth immersion of a domain of dimension 
$n-1$ in $M$, with a Riemannian induced metric. At 
any point $P\in\Sigma $, there is a timelike unit normal vector $N_P$,
$g(N_P,N_P)=-1$,
called the ``future pointing Gauss map'' of the hypersurface, with 
the orientation of $\partial_t(P)$ (hereafter we omit to specify $P$). 
The normal vector and the tangent space of $\Sigma $ 
at $P$ provide the decomposition $\partial_t=\alpha N + Y$, 
where $\alpha >0$, $g(N,Y)=0$. 
From $-1=g(\partial_t,\partial_t)=-\alpha^2+g(Y,Y)$ it follows that $\alpha \ge 1$; the value 
$\alpha =-g(N,\partial_t) =\cosh\theta $ defines the normal hyperbolic angle $\theta $ 
of the hypersurface at 
$P$\footnote{At a point $P\in \Sigma $, the value $\cosh \theta$ is the 
{\em energy} of the observer $N$ measured by the istantaneous comoving observer 
$\partial_t$, while $N  \tanh\theta $ is the velocity \cite{SacWu}.}. 
The tangential component $Y$ of the decomposition %
introduces the ``height function" $h(P)$
of the hypersurface through the relation $Y=\nabla h$. It is 
$|\nabla h|^2 = \sinh^2\theta $.\\
Given a parametrization $x^i=x^i(\vec q)$ of $\Sigma $, with basis tangent vectors $B^i_\mu = \partial x^i/\partial q^\mu$, it is $\nabla_\mu B^i_{\nu} = \nabla_\nu B^i_\mu $, $g_{ij}N^i B^j_\mu =0$, and the induced metric is
$(g_\Sigma)_{\mu\nu} = g_{ij} B^i_\mu B^j_\nu$. In the relation $ \nabla_\nu B^i_{\mu} = -N^i\Omega_{\mu\nu} $, the symmetric matrix $\Omega_{\mu\nu}$
is the ``second fundamental form" of the immersion (see \cite{LovRun}, eq.5.4). Its eigenvalues $k_1,\dots, k_{n-1}$
are the principal curvatures of $\Sigma $ at $P$. 
The mean curvature, the second mean curvature, etc., are  
$$H=\frac{1}{n-1}\sum_{j=1}^{n-1} k_j, \quad H_2 =  \frac{2}{(n-1)(n-2)} \sum_{i<j} k_i k_j, \quad \text{etc.}$$
A spacelike hypersurface is ``maximal'' if $H=0$, it is ``complete" if ($\Sigma, g_\Sigma $) is a complete Riemannian manifold (any geodesic starting from a point
in $\Sigma $ is isometric to the real line). A maximal spacelike hypersurface is the transition between expanding and recontracting phases of the universe. In Minkowski spacetime $\mathbb R^n_1$ the only complete maximal spacelike hypersurfaces are spacelike hyperplanes (Calabi, Cheng and Yau).\\
A point $P\in\Sigma $ is ``umbilical'' if $\Omega (P)$ is proportional to 
the metric tensor $g_{\Sigma}(P)$. The hypersurface is ``totally umbilical'' 
if every point of $\Sigma$ is umbilical. Photon surfaces in
general relativity are {\em timelike} totally umbilical submanifolds.

Slices $\Sigma_t=\{t \}\times M^*$ are particularly simple spacelike hypersurfaces. The hyperbolic angle of a slice is always $\theta=0$ and the 
inherited metric tensor is $g_t=f^2(t) g^*$. Since $B^0_\mu=0$ and $N^i =\delta_{i0}$ in the warped coordinate frame, 
one finds $\Omega_{\mu\nu} = \Gamma^0_{\mu\nu} = f^{\prime}f g^*_{\mu\nu}$. Therefore a slice is totally
umbilical. Being $\Omega^\mu_\mu = (f^\prime/f) (n-1)$, the mean curvature is $H=f^\prime/f$.

Constant mean curvature spacelike hypersurfaces are studied in geometry and in general relativity for the initial value problem of field equations. 
They are critical points of the area functional under certain volume constraints 
\cite{Barbosa}. A lower bound for their curvature scalar is derived in \cite{AleRub16}, 
with the assumption $(\log f)^{\prime\prime} \le 0$. If $N^*$ is the projection of the
unit normal vector $N$ of the hypersurface $\Sigma $ on the tangent space of $M^*$, it is: 
\begin{align}
R_\Sigma \ge \frac{R^*}{f^2} + 2R^*_{\mu\nu} N^{*\mu}N^{*\nu} +
(n-1)(n-2) [(f^\prime/f)^2 -H^2].
\end{align}
Equality holds if and only if  $\Sigma $ is totally umbilical and $(\log f)^{\prime\prime} 
\sinh^2 \theta =0$.

Are slices the only constant mean curvature spacelike hypersurfaces?
The answer to this problem has been connected with the following ``timelike convergence condition'' (TCC) (see \cite{[Ali3],SacWu}): 
\begin{align}\label{eq_30.2}
R_{ij} X^i X^j \ge 0 \quad \forall X: \; X^2<0
\end{align} 
It translates the fact that gravity, on average, is attractive.
For a GRW manifold, if the timelike eigenvector of the Ricci tensor is used, 
the TCC implies $\xi\le 0$ i.e.
$f^{\prime\prime}\le 0$ ($f$ is positive). For a general timelike vector $(X^0,V^\mu)$, 
$X^2 = - (X^0)^2 + f^2 g^*_{\mu\nu} V^\mu V^\nu$, the
TCC written in the components \eqref{eq_3.1} of the Ricci tensor is: 
$$
R^*_{\mu\nu}V^\mu V^\nu \ge  -(n-1)(f^{\prime\prime}/f) X^2   +(n-2) g^*_{\mu\nu}V^\mu V^\nu ( ff^{\prime\prime} - f^{\prime 2} )$$
for any value $X^2<0$ and any vector $V$ tangent to the fiber. As $f^{\prime\prime}\le 0$ and since $X^2$ can be as
small as wanted, a GRW spacetime satisfies the TCC property if and only if
\begin{gather}
R^*_{\mu\nu}V^\mu V^\nu \ge  (n-2) g^*_{\mu\nu}V^\mu V^\nu \sup_{t\in I} ( ff^{\prime\prime} - f^{\prime 2} ),\\
f^{\prime\prime}\le 0 
\end{gather}

\begin{thrm}[Al{\'\i}as \& al., 1995, \cite{AliRomSan95_a}] 
Let $M$ be a TCC GRW spacetime and $\Sigma $ a compact spacelike hypersurface of constant mean curvature $H$, then:\\
1) $\Sigma $ is totally umbilical; 2) either $\Sigma $ is a spacelike slice, or there exist a constant $c>0$ and $t_0\in I$ such that: 
i) $f(t)= \sqrt{c/c_1}  \sin (\sqrt {c_1}t  + c_2)$ or 
$f(t)= \sqrt c t + c_3$ in a neighborhood of $t_0$  for some real constants $c_1>0$ and $c_2$, $c_3$, ii) $Hf'>0$ on $\Sigma $, iii) $R_{ij}Y^iY^j \ge -c Y^2$ for all vectors
$Y$ (and there exists a nonvanishing vector $Z$ in an open set in $M$ where equality holds).
\end{thrm}

Montiel proved that the condition $f^{\prime\prime}\le 0$ alone, without
any hypothesis on $R_{ij}^*$, is sufficient for
the spacelike slices to be the only compact constant mean curvature spacelike hypersurfaces of the manifold.
The condition was weakened to requiring $ff^{\prime\prime}-f^{\prime 2}\le 0$ \cite{AleRub16}.

\begin{thrm}[Montiel, 1999, \cite{Montiel99}] 
Let $f : I \to \mathbb R$ be a positive smooth function defined on an open
interval, such that $ff^{\prime\prime}-f^{\prime 2}\le 0$. Then, the
only compact spacelike hypersurfaces immersed into a GRW spacetime and having constant mean curvature are the slices $\{t\} \times M^*$, for a (necessarily compact) Riemannian manifold $M^*$.
\end{thrm}

\begin{thrm}[Montiel, 1999, \cite{Montiel99} theorem 4] In a spacetime $M$ with a closed conformal timelike field 
$X$ (i.e. $u^k \nabla_k X^i=\varphi  u^i$ for any $u$) and obeying the null convergence condition ($R_{ij}v^iv^j\ge 0$ for all null directions), compact spacelike hypersurfaces $\Sigma $ with constant mean curvature must be umbilical.
\end{thrm} 
\noindent
Montiel observed  that every conformally stationary
spacetime admitting a timelike closed conformal vector field is locally
isometric to a GRW spacetime (for a global assertion see \cite{Montiel99}, Proposition 2).

A GRW spacetime $M$ is ``spatially closed'' if $M^*$ is compact; if $M$ admits a compact spacelike 
hypersurface, then it is spatially closed. Bounds on the volume of compact spacelike hypersurfaces are 
given in \cite{AlRomRub14}. \\
When the fiber $M^*$ is non-compact, no spacelike hypersurface 
can be compact \cite{AliRomSan95_a}. 
The problem of uniqueness of complete noncompact spacelike hypersurfaces is 
studied by Dong and Liu by bounding the ratio $H_2/H$ or a higher order one \cite{DongLiu15}.\\
By assuming that the 
hyperbolic angle $\theta $ of the hypersurface has a local maximum at a point $P_0$, and under certain curvature assumptions at the maximum point, Latorre and Romero proved that, locally, the hypersurface is a slice:
\begin{thrm}[Latorre \& Romero, 2002, \cite{LatRom02}]
Let $M$ be a GRW spacetime and let $\Sigma $ be a spacelike
hypersurface of constant mean curvature. If the hyperbolic angle $\theta $
of the hypersurface attains a local maximun at some point $P_0\in \Sigma$ and if\\
- either $f^{\prime\prime}<0$ at $t(P_0)$ and $R^*_{\mu\nu}$ is positive semi-definite at $P_0$, \\
- or $f^{\prime\prime}\le0$ at $t(P_0)$ and $R^*_{\mu\nu}$ is positive definite
at $P_0$, \\
then, there exists an open neighborhood of $P_0$ in $\Sigma $ which is a 
slice.
\end{thrm}

Geometric conditions for a constant mean curvature
hypersurface to be a slice of the foliation were investigated in \cite{AliRomSan97,AliMon01,CabRomRub10,AliRigSco15}. \\
In \cite{AliEst96} an inequality for the Gaussian curvature of compact maximal surfaces is found, that saturates for the totally geodesic ones. Bounds
for the mean curvature of spacelike graphs in GRW spacetimes are in \cite{AliImpRig12}. 
Hypersurfaces with constant higher order mean curvature are studied in \cite{AliCol07,AliImpRig12}, and the height function is studied in \cite{GarImp14}.

A hypersurface in a Lorentzian manifold $\Sigma $ is ``null''  if the induced metric tensor is
degenerate on it (at each point $P$ there is a tangent vector $\xi\neq 0 $ such that $g(\xi, Y)=0$ for all $Y\in T_P\Sigma $). The light-cone 
in Minkowski space, or black-hole horizons are examples. Null hypersurfaces have been studied by Navarro et al. \cite{NavPalSol16}, Duggal \cite{Duggal16}.
Gutierrez and Olea focused on totally umbilical null hypersurfaces in GRW manifolds. They proved that through each point of $M$ such surfaces appear in 
pairs, giving rise to a local decomposition of the fiber as a twisted product.
Nullcones are the unique totally umbilic null hypersurfaces in the closed Friedmann cosmological model \cite{GutOle15}.

Spacelike hypersurfaces for spatially parabolic GRW spacetimes (i.e. parabolic fiber $M^*$\footnote{A
noncompact complete Riemannian manifold such that the only 
superharmonic functions on it, which are bounded from below, are the constants.}) 
have been studied by Romero et al. Such spaces model open cosmologies, 
which are favoured by present observations. The main result is the following, which includes several prior results:
\begin{thrm}[Romero \& al., 2013, \cite{RomRubSal13}]
Let $\Sigma $ be a complete spacelike hypersurface in a GRW spacetime, 
whose fiber $M^*$ has parabolic universal Riemannian covering. If the 
hyperbolic angle of $\Sigma $ is bounded and the restriction to $\Sigma $ of the warping function $f$ satisfies $\sup f(t) <\infty $ and $\inf f(t)>0$, then, 
$\Sigma $ is parabolic.
\end{thrm} 
The problem of stability of the area  of constant mean curvature spacelike hypersurfaces under volume variations has been addressed by Barbosa and do Carmo (1984). We quote a recent result for GRW spacetimes:
\begin{thrm}[Barros \& al., 2008, \cite{Barros16}]
Let $\Sigma $ be a closed spacelike hypersurface of a GRW spacetime, 
with constant mean curvature $H$. If the warping function $f$ satisfies $f^{\prime\prime}\ge \max \{
H f^\prime ,0\} $ and $\Sigma $ is strongly stable, then $\Sigma $ is either maximal or a spacelike 
slice $\{ t_0 \} \times M^*$, for some $t_0 \in I$.
\end{thrm}
The index form along time-like geodesics on a Lorentzian warped manifolds  is studied in \cite{EhrKim05} and applied to GRW spacetime.

Gutierrez and Olea \cite{GutOle09} studied the conditions on the curvature of a Lorentzian manifold to achieve 
a global decomposition as a GRW spacetime. While the existence of a timelike closed conformal vector field implies that
a Lorentzian manifold is locally a GRW spacetime \cite{Sanchez98}, it does not 
imply that the decomposition is global.\\
They proved that de Sitter spaces are the only nontrivial complete
Lorentzian manifolds with more than one GRW decomposition, while Friedmann cosmological models 
admit a unique GRW decomposition, even locally.\\
They identify a reference frame with a timelike unit vector field $u_j$. If it is closed, orthogonally conformal, 
and such that $\nabla_i\nabla^j u_j$ is proportional to $u_i$, it is named a ``warped reference frame''.
\begin{thrm}[Gutierrez \& Olea, 2009, \cite{GutOle09}, proposition 2.2]
Let $I$ be an open real interval, $(M^*,g^*)$ a manifold and $g$ a Lorentzian metric on $I\times M^*$ such that the canonical
foliations are orthogonal. If $u=\partial_t$ is a warped reference frame,
then $g$ is the warped product $g_{ij}dx^idx^j=-(dt)^2 + f(t)^2 g^*_{\mu\nu} dx^\mu dx^\nu$ and 
$f(t) = \exp \left [ \frac{1}{n-1} \int_0^t ds \nabla^j u_j (s,x) \right ]$, being $x$ a point of $M^*$.
\end{thrm}
\begin{thrm}[Gutierrez \& Olea, 2009, \cite{GutOle09}, theorem 3.2]
Let $M$ be a complete non-compact Lorentzian manifold with $n\ge 3$ and $u$ a non-parallel warped
reference frame, and let $R_{ij}$ be the Ricci tensor. If one of the following conditions is true: 1) $R_{ij} u^i u^j\le 0$, 2)
$R_{ij}\nu^i\nu^j \ge 0$ for all $\nu\perp u$, 3) $R_{ij}w^iw^j \ge 0$ for all light-like vectors $w$, 
then $M$ is globally a GRW spacetime.
\end{thrm}
A more refined result was obtained by Caballero et al.  
\begin{thrm}[Caballero \& al., 2011, \cite{CabRomRub11}, theorem 3.1]
A Lorentzian manifold admits a global decomposition as a GRW spacetime if 
it is a spacetime equipped
with a timelike gradient conformal vector field $X_j$, such that the flow of the vector field
$u_j = X_j/ \sqrt{-X^2}$ is well defined and onto a domain $I\times L$, for some real interval $I$
and some leaf $L$ of the foliation orthogonal to $X$.
\end{thrm}
\section{\bf Killing vectors}
Local and global characterizations of GRW spacetimes in terms of  
Killing or conformal Killing vectors are given in \cite{Sanchez98,Sanchez99,GutOle09,CabRomRub11}. 
\begin{definition}[\cite{Stephani}]
A vector field is named {\em conformal Killing} if $\nabla_k X_j + \nabla_j X_k = \theta g_{jk}$, being $\theta $ a scalar function. It is named Killing if $\theta=0$, homothetic if $\theta=$const.\\
If the conformal Killing vector is closed, $\nabla_k X_j=\nabla_j X_k$, 
then it is concircular in the sense of Fialkow (see def.\ref{Fialkow}).
\end{definition}
A conformal change of the metric of a GRW spacetime with a conformal factor that only depends on $t$, 
produces a new GRW spacetime \cite{[Ali3]}.\\
With $dt' = dt/f(t)$, the metric \eqref{eq_1.1} becomes conformally equivalent 
to $-dt^{\prime 2}+g^*_{\mu\nu}dx^\mu dx^\nu $. S\'anchez \cite{Sanchez98} concluded  
that the vector $X=f\partial_t$ is parallel in the new metric,  $\nabla^\prime X =0$, and 
is conformal Killing in any metric conformal to it. In particular in the metric \eqref{eq_1.1} it is
$\nabla_k X_j +\nabla_j X_k = 2 f' g_{kj}$. \\
He then enumerated the special properties of the vector field $u=\partial_t$. It is:
i) timelike unit;  ii) geodesic  ($u^i\nabla_i u_j=0$); iii) spatially conformal 
($\nabla_\mu u_\nu + \nabla_\nu u_\mu = 2(f^\prime/f) g_{\mu\nu}$);  
iv) irrotational ($\nabla_i u_j-\nabla_j u_i=0$); v) $\nabla_i\nabla^j u_j$  is pointwise parallel to $u_i$. 
In ref.\cite{Sanchez99} he proved that it is an eigenvector of the 
Ricci tensor. \\
The following theorem was achieved:
\begin{thrm}[S\'anchez, 1998, \cite{Sanchez98}, theorem 2.1]\label{thrm_2.1}
Let $(M,g)$ be a simply connected Lorentzian manifold with a complete vector field $u$
(i.e. its flow curves exist for all $t$) satisfying items (i)-(v), then 
$(M,g)$ is (globally) a GRW spacetime with $u=\partial_t$.
\end{thrm}

A spacetime that admits a timelike Killing vector field $X$ is called stationary; 
a set of coordinates $(t,x^\mu)$ can be found such that $X=\partial_t$,
$\partial_\mu$ are spacelike, and the metric tensor does not depend on $t$ \cite{[Ali3]}. If 
$X$ is also irrotational, the spacetime is called static.\\
A Lorentzian manifold with a timelike conformal Killing vector field $X$, can be
conformally mapped to a Lorentzian manifold where $X$ is a timelike Killing vector 
field \cite{[Ali3]}. Spacetimes with conformal Killing tensors were
investigated in \cite{DafDad94}. They are special cases of ``conformally stationary spacetimes", which allow for an observer seeing an isotropic microwave background 
\cite{Fer}. A classical result  \cite{[Ehl68]} assures that RW spacetimes are characterized as the spacetimes that 
admit a geodesic observer who sees an isotropic microwave background and whose stress-energy 
tensor is a perfect fluid comoving with the observer. The latter property cannot  strictly hold in cosmological 
models describing the formation of structures. 
In this way GRW spacetimes
are a privileged class of inhomogeneous spacetimes admitting an isotropic radiation \cite{Sanchez98}.\\ 
If $X_j$ is globally the gradient of some smooth function, it is called ``gradient conformal" vector
field \cite{CabRomRub11}. In this case the spacetime admits a global time function and it is stably 
causal \cite{Hawking68},
i.e. there is a finite neighbourhood of the original metric of the spacetime such that any of its Lorentzian metrics is causal \cite{[Bee96],CabRomRub11}. 
We refer also to the works of Romero et al. \cite{RomRubSal13,[Rom2]} and Guti\'errez and Olea \cite{GutOle09} for a presentation of geometric and physical properties.

S\'anchez studied some global properties of GRW spacetimes, focusing on geodesics \cite{Sanchez98} and conformal Killing vectors \cite{Sanchez99}. 
If a GRW spacetime admits a non-trivial Killing vector, then the warping function $f$ must be one listed in Tables 1, 2 of \cite{Sanchez99}. 
%
%%%%%%%%%%%%%%%%%%%%%%%%%%%%%%%%%%%%%%%%%%%%%%
%
\section{\bf Chen's theorem}

A turning point in the literature was the characterization of GRW spacetimes by the presence of a timelike concircular vector field, by Bang-Yen Chen in 2014 (theorem \ref{thrm_2.7}). We show some simple but significative properties of the Ricci and the Weyl tensors of a pseudo-Riemannian manifold with a concircular tensor
(theorem \ref{thrm_3.7}). Such properties belong to all GRW spacetimes.\\
A characterization in terms of a torse-forming vector field descends from Chen's theorem (theorem \ref{thrm_2.8}).
Another one, in terms of a conformal Killing tensor, will be given in the next section. 
\begin{definition}
A vector field is named {\em torse-forming} if $\nabla_k X_j =\omega_k X_j +\varphi g_{kj}$, being $\varphi $ a scalar function and $\omega_k$ a non
vanishing 1-form. 
\end{definition}
\noindent
Riemannian spaces with a torse-forming vector field were studied by Yano as early as 1944 \cite{Yano40,Yano44}. The study was extended to pseudo-Riemannian spaces
by Sinyukov \cite{Siny79}, Mike\v{s} and Rach\r{u}nek \cite{MikRach,RachMik05}. 
Note that for a unit timelike ($u^2=-1$) vector, the torse-forming condition becomes $\nabla_k u_j = \varphi (u_k u_j +g_{kj})$.\\
The presence of a torse-forming vector implies the following shape of the
metric: 
\begin{align}
ds^2 = \pm (dt)^2 + F(t,\vec x) d s^{*2}  \label{eq_warped_Yano}
\end{align}
 where $ds^{*2}$ is the metric of 
the submanifold parametrized by $\vec x$. 

Torse-forming vectors associated to 1-forms $\omega_k$ that are locally the gradient of a 
scalar function were named concircular by Yano. For them $F$ is only a function of $t$. 
Concircular vectors appeared
in the study of conformal mappings preserving geodesic circles \cite{Yano40,Yano44}, and in the theory of projective and conformal transformations. In de Sitter's model of general relativity, 
the world lines of receding or colliding galaxies are trajectories of timelike 
concircular vector fields \cite{Takeno67}.\\
Fialkow \cite{Fialkow39} gave a definition different from Yano's: 
\begin{definition}\label{Fialkow}
A vector field $X$ is {\em concircular} if, for a scalar function $\rho $, it is 
\begin{align}\label{eq_1.2}
\nabla_k X_j = \rho g_{kj}
\end{align}
%being $\rho $ a scalar function. 
\end{definition}
In this paper we adopt Fialkow's definition.
The existence of a concircular vector (not necessarily timelike) on a pseudo-Riemannian 
manifold, hence all GRW spacetimes, poses restrictions on the curvature
tensors:
\begin{thrm}[Mantica \& Molinari, 2016, \cite{ManMol16a}]\label{thrm_3.7}
If a pseudo-Riemannian manifold is equipped with a concircular vector,  $\nabla_j X_k = \rho g_{jk}$, 
$X^2\neq 0$, then:\\
$X$ is an eigenvector of the Ricci tensor:
\begin{align}
&R_{im}X^m = \xi X_i , \qquad \xi = -(n-1) \frac{X^m\nabla_m \rho}{X^2}, \label{eq_RicX}\\
& \nabla_i \xi = X_i\theta \label{eq_theta}
\end{align}
where $\theta $ is a scalar function.
\begin{align}\label{eq_RiemannX}
R_{ijk}{}^m X_m = - \frac{\xi}{n-1}(X_i g_{jk} - X_j g_{ik} )
\end{align}
The Ricci tensor has the general expression, involving the Weyl tensor:
\begin{align}\label{eq_3.10}
R_{kl} = \frac{n\xi -R}{n-1} \frac{X_kX_l}{X^2} - \frac{\xi-R}{n-1}g_{kl}  
+(n-2) C_{aklb}\frac{X^aX^b}{X^2}
\end{align}
The vector $X$ is Riemann and Weyl compatible \cite{ManMol12b}:
\begin{align}
X_iX^m R_{jklm} + X_jX^m R_{kilm} + X_kX^m R_{ijlm}=0, \label{eq_Rcomp}\\
X_iX^m C_{jklm} + X_jX^m C_{kilm} + X_kX^m C_{ijlm}=0. \label{eq_Wcomp}
\end{align}
\end{thrm}
\quad\\
Some useful consequences are listed:\\
1) A covariant derivative of $R_{ij}X^j=\xi X_i$ gives
\begin{align}\label{eq_3.14}
X^j\nabla_k R_{jl} = -\rho R_{kl} +\theta X_kX_l +\rho \xi g_{kl}
\end{align}
with its contracted form $X^m\nabla_m R = 2n\rho\xi -2\rho R + 2X^2\theta $.\\
2) The property of Weyl compatibility \eqref{eq_Wcomp} gives:
\begin{align}\label{eq_3.15}
C_{jklm}X^m = (X_j C_{aklb} -X_k C_{ajlb}) \frac{X^aX^b}{X^2}.
\end{align}
It implies that $C_{jklm}X^m=0$ if and only if $C_{jklm}X^jX^m=0$.\\
3) The relation $R_{im}X^m = - (n-1)\nabla_i \rho $ shows the useful identity:
\begin{align}
\nabla_k \rho = -\frac{\xi}{n-1} X_k \label{eq_nablarho}
\end{align} 
\begin{thrm}[Chen, 2014, \cite{Chen14}]\label{thrm_2.7}
A Lorentzian manifold $M$ of dimension $n\ge 3$ is a GRW spacetime if and only if it 
admits a timelike concircular vector: $X^2<0$ and 
$\nabla_k X_j = \rho g_{kj}$.
\end{thrm}
Besides its simplicity, Chen's theorem is a covariant
characterization of GRW spacetimes. Hereafter, we shall refer to the timelike concircular vector field $X$ as to ``Chen's vector''.\\
Some points are worth remarking.
If $M$ is a GRW spacetime, then there is a warping coordinate frame \eqref{eq_1.1}.
The timelike vector $X= \lambda(t)\partial_t$ is an eigenvector of the Ricci tensor with eigenvalue \eqref{eq_xi1} for any function $\lambda $. 
With the Christoffel symbols \eqref{eq_Christoffel}, the condition $\nabla_i X^j = \rho \delta_i{}^j$ becomes $\partial_t \lambda (t) = \rho$. 
By theorem~\ref{thrm_3.7}, Chen's vector is eigenvector of the Ricci tensor with eigenvalue \eqref{eq_RicX}, then $\lambda (t)=f(t)$ and $\rho (t) = f^\prime (t)$, where $f$ is the warping function. 

\begin{remark}
The Weyl compatibility of Chen's vector \eqref{eq_Wcomp} 
implies that GRW spacetimes are purely electric, 
and the algebraic types of the Weyl tensor can only be 
$G$, $I_i$ $D(d)$ or $O$ {\rm (}\cite{HerOrtWyl12}, proposition 4.10{\rm )}.
As a matter of fact, in \cite{HerOrtWyl12} it is shown that a spacetime 
with $ds^2 = -V^2 (\vec x , t) dt^2 + P^2 (\vec x , t)\tilde g_{\mu\nu} (\vec x)
dx^\mu dx^\nu$, of which \eqref{eq_1.1} is a particular
case, is purely electric. 
\end{remark}

\begin{remark}
By evaluating the Ricci tensor \eqref{eq_3.10} in the warped frame 
\eqref{eq_1.1} where Chen's vector has components $X^0=f$ and $X^\mu=0$, and  comparing it with
the expression \eqref{eq_3.1}, the following relation is found: 
$$ R^*_{\mu\nu} - \frac{R^*}{n-1}g^*_{\mu\nu} = -(n-2) C_{0\mu\nu0} $$  
By means of \eqref{eq_3.15} one evaluates $C_{\mu \nu \rho 0} = 0$.
\end{remark}

Given a timelike concircular vector $X$, the unit
timelike vector $u_k = X_k/\sqrt {-X^2}$ is torse-forming: 
\begin{align}\label{nablaN}
\nabla_j u_k = \frac{\rho}{\sqrt {-X^2}}(g_{jk}+u_j u_k)\equiv \frac{\rho}{\sqrt {-X^2}} h_{jk}
\end{align}
where $h_{ij}$ is a projector on vectors perpendicular to $u$ (the tangent space of a fiber). Being $u^j\nabla_j u_k =0$, the integral curves of $u_j$ are geodesics. \\
The torse-forming property is weaker than concircularity, but with an additional condition, the following characterization of GRW spacetimes is possible:
\begin{prop}[Mantica \& Molinari]\label{thrm_2.8}
A Lorentzian manifold of dimension $n\ge 3$  is a GRW spacetime if and only if 
it admits a unit timelike torse-forming vector, $\nabla_k u_j = \varphi (g_{kj} + u_k u_j)$, 
that is also an eigenvector of the Ricci tensor.
\begin{proof}
The integrability conditions of $\nabla_k u_j =\varphi (u_ju_k+g_{jk})$ is
$$ R_{jkl}{}^m u_m = u_ku_l \nabla_j \varphi -  u_ju_l \nabla_k \varphi + \varphi^2(u_k g_{jl}-u_j g_{kl}) + g_{kl} \nabla_j \varphi - g_{jl}\nabla_k \varphi $$
from which it is $R_j{}^m u_m = (2-n)\nabla_j \varphi + u_ju^l\nabla_l \varphi + \varphi^2(n-1) u_j $. If $u_j$ is an eigenvector of the Ricci tensor, then $\nabla_j \varphi =Fu_j$
for some scalar $F$. Then, $\nabla_j( \varphi u_k)-\nabla_k(\varphi u_j) = 0$ i.e. $\varphi u_k$ is locally a
gradient: $\varphi u_k=\nabla_k \sigma$. On defining $X_l=u_l e^{-\sigma}$
we have $X^2<0$ and $\nabla_kX_l= (\varphi e^{-\sigma}) g_{kl}$. Thus the spacetime is GRW.\\
The opposite way, if a Lorentzian manifold is GRW, then there exists a timelike concircular vector $\nabla_k X_j=\rho g_{kj}$. 
Then $u_j = X_j /\sqrt{-X^2}$ is torse-forming, $u^2=-1$ and, by theorem \ref{thrm_3.7}, $u_j$ is an eigenvector of the Ricci tensor.
\end{proof}
\end{prop}

\section{\bf GRW spacetimes and conformal Killing tensors.}
We give a new characterization of GRW spacetimes based on a conformal Killing tensor of special form. 
\begin{definition}[see \cite{RanEdgBar03}]
A conformal Killing tensor is a symmetric tensor $K_{ij}$ satisfying the condition
\begin{align}\label{eq_1.3} 
\nabla_i K_{jk} + \nabla_j K_{ki} + \nabla_k K_{ij} = \eta_i g_{jk} + \eta_j g_{ki} +\eta_k g_{ij}
\end{align}
being $\eta_i$ a non-vanishing 1-form called associated conformal vector.
\end{definition}
\noindent
If $\eta_i$ is a Killing vector, $K_{ij}$ is a ``homothetic Killing tensor" \cite{RanEdgBar03};
if $\eta_i $ is the gradient of a scalar function $\eta $ then $K_{ij}$ 
is a ``gradient conformal Killing tensor", and $K_{ij} - \eta g_{ij}$ is a Killing tensor 
called ``associated Killing tensor".\\ 
Symmetric Killing tensors define first integrals of the equations of motions, i.e. functions
which are constant on geodesics; conformal Killing tensors define first integrals for null geodesics. In 1970
Walker and Penrose obtained  a conformal Killing tensor as the first integral of the null
geodesic equations of every type of $\{2,2\}$ vacuum solutions of Einstein's equations in $n=4$ \cite{WalPen70}. In the same
paper, the charged Kerr solution was shown to admit a Killing tensor which, together with the metric and
the two Killing vectors, allowed for the explicit integration of the geodesics. Killing and conformal Killing
tensors appear in the study of the geometric inverse problems, integrable systems, Einstein-Weyl geometry
(see \cite{HeMoSe16} and references therein).
\begin{lem}[Mantica \& Molinari]\label{prop_2.10}
Let $K_{ij}$ be a conformal Killing tensor with an eigenvector $u^2\neq 0$:
$K_{jk}u^k = \lambda u_j$. If $u^m\nabla_m u_k = \theta u_k$ for some (possibly zero) scalar $\theta $, then $\eta_i =\nabla_i\lambda $.
\begin{proof}
A covariant derivative gives $u^k \nabla_i K_{jk} + K_{jk} \nabla_i u^k = u_j\nabla_i\lambda
+\lambda \nabla_i u_j$. Contraction with $u^j$ and $u^i$ give: 
$ u^j u^k \nabla_i K_{jk} = u^2 \nabla_i\lambda$ and
$$ u^ku^i\nabla_i K_{jk} = - K_{jk} u^i \nabla_i u^k + u_j u^i \nabla_i \lambda + \lambda u^i\nabla_i u_j $$
Multiplication of \eqref{eq_1.3} by $u^j u^k$ and the use of the above relations give
$$ u^2 \nabla_i\lambda - 2(K_{il}-\lambda g_{il})u^m \nabla_m u^l  + 2u_i u^m\nabla_m\lambda
= u^2\eta_i + 2u_i u^k\eta_k $$
Contraction with $u^i$ gives $u^2 u^i (\eta_i -\nabla_i\lambda)=0$. If $u^2\neq 0$ the equation simplifies
\begin{align}\label{eq_2.9}
u^2(\nabla_i \lambda -\eta_i) =2 (K_{il}-\lambda g_{il}) u^m \nabla_m u^l 
\end{align}
and if $u^m\nabla_m u^l=\theta u^l$ the result follows.
\end{proof}
\end{lem}
\begin{thrm}[Mantica \& Molinari]\label{thrm_2.12}
A Lorentzian manifold is a GRW spacetime if and only if there is a conformal Killing tensor of the type
\begin{align}\label{eq_2.12}
K_{ij} = A g_{ij} + B u_i u_j
\end{align}
where $A$, $B$ are scalar fields, $B\neq 0$, $u^2=-1$ and $\nabla_k u_j =\nabla_j u_k$.
\begin{proof}
Suppose that $K_{ij}$ is a conformal Killing tensor with the form \eqref{eq_2.12}. Since $K_{ij}u^j = (A-B)u_i$ and $u^k\nabla_k u_j=0$,
by Lemma \ref{prop_2.10} it is $\eta_i = \nabla_i(A-B)$. The property that
$K_{ij}$ is conformal Killing becomes 
\begin{align}\label{eq_2.14}
& B[\nabla_j (u_ku_l) + \nabla_k (u_ju_l) + \nabla_l (u_ku_j)]\\
&=  -(g_{kl}+u_ku_l)\nabla_j B - (g_{jl}+u_ju_l)\nabla_k B - (g_{jk}+u_ju_k)\nabla_l B
\nonumber
\end{align}
A contraction with $u^l$ and the closedness property give 
\begin{align}\label{eq_2.15}
\nabla_j u_k = (g_{jk}+u_ju_k) \frac{u^m\nabla_m B}{2B}
\end{align}
The equation proves that $u_k$ is torse-forming. It is used to simplify \eqref{eq_2.14}: 
\begin{align*} & u^m \nabla_m B (3u_iu_ju_k +g_{ij}u_k + g_{jk} u_i + g_{ik}u_j) = \\
&=  -(g_{kl}+u_ku_l)\nabla_j B - (g_{jl}+u_ju_l)\nabla_k B - (g_{jk}+u_ju_k)\nabla_l B.
\end{align*}
Contraction with $g^{ij}$ gives: $u_k u^m\nabla_m B = -\nabla_k B$. The torse-forming
property \eqref{eq_2.15} becomes:
\begin{align}\label{eq_2.17}
\nabla_j u_k = \frac{u^m\nabla_m B}{2B} g_{jk} - \frac{\nabla_jB}{2B} u_k
\end{align}
The vector $X_k= e^\sigma  u_k$ is timelike and concircular with the choice
$\nabla_j\sigma = \frac{1}{2}(\nabla_j B)/B$: 
$$\nabla_j X_k = \left[e^\sigma \frac{u^m\nabla_m B}{2B}\right ] g_{jk}. $$ 
By Chen's theorem \ref{thrm_2.7} the manifold is a GRW.\\
The converse is easily proven: let the spacetime to be a GRW. By theorem \ref{thrm_2.7} there exists a timelike concircular vector 
field $X_k$. Direct evaluation shows that 
\begin{align}\label{eq_2.20}
K_{ij} = (\rho - X^2)g_{ij}+X_iX_j 
\end{align}
is a conformal Killing tensor, with $\eta_i = \nabla_i\rho$. Next, define 
$u_i=X_i/\sqrt{-X^2}$; it is $u^2=-1$, $\nabla_j u_k = \nabla_k u_j$ and $K_{ij} = (\rho - X^2) g_{ij} + (-X^2) u_iu_j$.
\end{proof}
\end{thrm}
%
%%%%%%%%%%%%%%%%%%%%%%%%%%%%%%%%
%
\section{\bf  Perfect fluid GRW spacetimes}
In this section we present conditions for GRW spacetimes to be perfect fluid spacetimes.
We also prove an important equivalence.
\begin{definition}\label{def_3.2}
A Lorentzian manifold is named ``perfect fluid spacetime" if the Ricci tensor has the form
\begin{align}
R_{ij} = Ag_{ij}+Bu_iu_j \label{eq_3.2}
\end{align}
where $A$ and $B$ are scalar fields and $u^2=-1$.
\end{definition}
Geometers identify the special form \eqref{eq_3.2} of the Ricci tensor as the defining property of ``quasi-Einstein manifolds"
(with arbitrary metric signature). The Riemannian ones were introduced by Defever and Deszcz in 1991
\cite{DefDes91} (see also \cite{ChaMai00}). Pseudo-Riemannian ones  
arose in investigating exact solutions of Einstein's equations \cite{DesHotSen01}. Robertson-Walker spacetimes are quasi-Einstein \cite{ONeill}.
\begin{thrm}[S\'anchez, 1999, \cite{Sanchez99}] \label{thrm_3.3}
A GRW spacetime $M$ is a perfect fluid if and only if $M^*$ is an Einstein manifold, 
i.e. $R^*_{\mu\nu} = \frac{R^*}{n-1}g^*_{\mu\nu}$,
\end{thrm}
\begin{thrm}[G\c{e}barowski, 1994, \cite{Gebar94a,Gebar94b}]\label{thrm_3.5}
For a warped product \eqref{eq_1.1} the fibers are Einstein, i.e. $R^*_{\mu\nu}=\frac{R^*}{n-1}
g^*_{\mu\nu}$, if and only if $\nabla_m C_{jkl}{}^m=0$. 
\end{thrm}
The theorems by S\'anchez and G\c{e}barowski together imply that a
GRW spacetime is a perfect fluid if and only if $\nabla_m C_{jkl}{}^m=0$. However, we have shown that
an equivalent algebraic condition can be given:
\begin{thrm}[Mantica \& Molinari, 2016, \cite{ManMol16a}] \label{thrm_3.9}
On  every GRW spacetime with Chen's vector $X_j$
\begin{align}\label{equivalence}
 \nabla_m C_{jkl}{}^m=0\quad \Longleftrightarrow \quad X_m C_{jkl}{}^m=0.
 \end{align}
 \end{thrm}
\noindent
The proof rests on Chen's characterization of GRW spacetimes by
means of a concircular vector.

The above results are here summarized:
\begin{thrm} \label{thrm_3.11}
On every GRW spacetime $M$ with Chen's vector $X_j$, the following statements are equivalent: 
\begin{align*}
 a) \quad & X^m C_{jklm}=0,\\
 b) \quad & \nabla_m C_{jkl}{}^m=0,\\
 c) \quad & R_{jk}=A g_{jk} + B\frac{X_jX_k}{X^2},\\
 d) \quad & R^*_{\mu\nu}= \frac{R^*}{n-1}g^*_{\mu\nu}.
\end{align*}
for suitable scalar fields $A$, $B$. By \eqref{eq_nablarho} it is 
\begin{align}
 \nabla_k \rho = -\frac{A-B}{n-1} X_k \label{eq_nablarho_pf}
 \end{align}
\end{thrm} 
Einstein's field equations \eqref{eq_Einstein} link the perfect fluid structure of the Ricci tensor to a perfect fluid
stress-energy tensor
\begin{align}\label{eq_pf}
T_{ij}=(\mu+p) u_i u_j +p g_{ij}
\end{align} 
where $u_j$ is the fluid's flow velocity, $u^2=-1$,  $p$ is the isotropic pressure and $\mu $ is the energy
density.  Then, the Ricci tensor may be expressed in terms of $p$ and $\mu$:
\begin{align*}
R_{ij} = 8\pi (\mu+p)u_iu_j + 8\pi\, \frac{\mu -p}{n-2} \,g_{ij} %\label{eq_4.2}
\end{align*}
Comparison with \eqref{eq_3.1} gives the following relations 
between the warping function $f$, the pressure and the energy density in a GRW perfect fluid spacetime
(the first one is the eigenvalue $\xi $ of the Ricci tensor, the second one descends from the scalar $R$):
\begin{align}\label{eq_4.444}
(n-1)\frac{f^{\prime\prime}}{f} = -8\pi \left [\frac{n-1}{n-2} p+\frac{n-3}{n-2} \mu\right ], \quad 16\pi \mu f^2 = R^* + (n-1)(n-2)f^{\prime 2}
\end{align}
A derivative in the second relation eliminates $R^*$. By further eliminating the warping function, they give the following equation of state ($\mu+ p \neq 0$):
\begin{align}\label{eq_4.445}
\left[ \frac{\mu'}{\mu+p}\right ]' = \frac{1}{n-2}[(n-1)p+(n-3)\mu ] +\frac{1}{n-1}\left[ \frac{\mu'}{\mu+p}\right ]^2 
\end{align}
%
%%%%%%%%%%%%%%%%%%%%%%%%%%%%%%%%%%%%%%%%%%%%%%%%%%%%%%%%%%%%%
%
\section{\bf Further characterizations of GRW spacetimes}
In this section we present recent characterizations of perfect fluid spacetimes as GRW
spacetimes, in arbitrary dimension.
 
De and Ghosh \cite{DeGho00} showed that if a perfect fluid
spacetime with closed $u_j$ is conformally flat then $u_j$ is a concircular vector. 
The result was extended
by Mantica and Suh to pseudo-Z-symmetric spacetimes \cite{ManSuh12a} and to weakly Z-symmetric spaces \cite{ManMol12a}. Recently we proved
\begin{lem}
Let $M$ be a Lorentzian manifold of dimension $n\ge 3$, with Ricci tensor
$R_{kl} =A g_{kl} + Bu_ku_l$, where $A$, $B$ are scalar fields, $B\neq 0$, and $u_k$ 
is a unit timelike vector, $u^2 =-1$. If $\nabla_m C_{jkl}{}^m=0$, then:
\begin{gather}
u_k\nabla_j \gamma - u_j \nabla_k \gamma = 0 \label{eq_5.11} \\
u_k \nabla_j B - u_j\nabla_k B =  B (\nabla_j u_k -\nabla_k u_j) \label{eq_5.22}
\end{gather}
where $\gamma =(n-2)A+B$.
\begin{proof}
The condition $\nabla_m C_{jkl}{}^m=0$ 
with the explicit form of the Ricci tensor, becomes: 
\begin{align}\label{eq_5.44}
\nabla_k(Bu_ju_l)-\nabla_j (Bu_ku_l)=-\tfrac{1}{2(n-1)}(g_{jl}\nabla_k\gamma -g_{kl}\nabla_j\gamma)
\end{align}
By transvecting with $g^{jl}$ it is: 
$(\nabla_k + u_k u^l\nabla_l) + B\nabla_l(u_ku^l ) =\tfrac{1}{2}\nabla_k\gamma $.
Multiply by $u_m$ and take the antisymmetric part:
\begin{align}
(u_m\nabla_k -u_k\nabla_m)B + B[u_mu^l\nabla_lu_k-u_ku^l\nabla_l u_m]
= \tfrac{1}{2}(u_m\nabla_k -u_k\nabla_m)\gamma \label{eq_55}
\end{align}
By transvecting \eqref{eq_5.44} with $u^ju^l$ it is $
(\nabla_k +u_ku^l \nabla_l)B+Bu^l\nabla_l u_k=\tfrac{1}{2(n-1)}(\nabla_k -u_ku^l\nabla_l)\gamma $. Multiply by $u_m$ and antisymmetrize:
\begin{align*} 
(u_m\nabla_k - u_k\nabla_m)B + B[u_mu^l\nabla_l u_k - u_ku^l\nabla_l u_m] = 
\tfrac{1}{2(n-1)}(u_m\nabla_k -u_k\nabla_m)\gamma
\end{align*}
This equation and \eqref{eq_55} imply the first assertion. Contraction of
\eqref{eq_5.44} with $u^l$ gives the second assertion.
\end{proof}
\end{lem}
\begin{remark}\label{rem_555}
The lemma implies that if a combination $C=\alpha A + \beta B$, where
$\alpha $ and $\beta $ are numbers, has the property $u_j\nabla_k C = u_k\nabla_j C$, 
then $u$ is closed, i.e. $\nabla_k u_j - \nabla_j u_k=0$. In particular, if the scalar curvature
$R=nA-B$ is a constant, then $u$ is closed. More generally, $u$ is closed if there exists
a differentiable local relation $F(A,B)=0$.
\end{remark}
\begin{thrm}[Mantica \& al., 2016,  \cite{ManMolDe16} theorem 2.1]\label{th_5.1}
Let $M$ be a Lorentzian manifold of dimension $n\ge 3$, with Ricci tensor
$R_{kl} =A g_{kl} + Bu_ku_l$, where $A$, $B$ are scalar fields, $B\neq 0$, and $u_k$ 
is a unit timelike vector, $u^2 =-1$. 
If $\nabla_m C_{jkl}{}^m=0$ and $\nabla_k u_j = \nabla_j u_k$, then 
$u$ is rescalable to a timelike concircular vector $X$, and $M$ is a GRW spacetime, with $ u^m C_{jklm}=0$.
\begin{proof}
On multiplying \eqref{eq_5.44} with $u_j$ and using the results \eqref{eq_5.11}, \eqref{eq_5.22} and the closedness
condition, we obtain
\begin{align}
\nabla_j u_k = - \frac{u^m\nabla_m \gamma}{2(n-1)B} (u_ju_k+g_{jk}). \label{eqnablau}
\end{align}
Since $u_j$ is an eigenvector of the Ricci tensor, in view of Proposition \ref{thrm_2.8} the spacetime is GRW.
Moreover, by the same Proposition, the vector $X_k=e^{\sigma}u_k$ is time-like concircular. By Theorem \ref{thrm_3.9}
it is also $u^m C_{jklm}=0$.
\end{proof}
\end{thrm}
Because of Remark \ref{rem_555}, the condition of closedness of the velocity field may be replaced by 
an equation of state, by constancy of the curvature scalar, or $A=0$ (this case $R_{kl}=B u_k u_l$ was studied in \cite{ManSuhDe16}). 
\begin{prop}[Mantica \& al., \cite{ManMolDe16} prop. 3.1] \label{prop_5.4}
A perfect fluid spacetime of dimension $n\ge 4$, with differentiable equation of state $p=p(\mu)$, $p+\mu\neq 0$,
and  $\nabla_m C_{jkl}{}^m=0$, is a GRW spacetime. The velocity vector field
is irrotational ($\nabla_j u_i = \nabla_i u_j$), geodesic ($u^k\nabla_k u_j =0$) and annihilates the Weyl tensor $u_m C_{jkl}{}^m=0$.
\end{prop}

In ref. \cite{Deszcz90} (lemma 4.1, theorem 4.1 and corollary 4.1) R. Deszcz
proved that a quasi-Einstein Riemannian manifold with harmonic Weyl tensor is, under certain conditions,
a warped product $I\times f^2 M^*$, where $M^*$ a Riemannian manifold of constant curvature.

Let's consider the case where the Ricci tensor has the perfect fluid form \eqref{eq_3.2} 
and is also conformal Killing,
\begin{align}\label{eq_RicciK}
\nabla_i R_{jk} + \nabla_j R_{ki} + \nabla_k R_{ij} = \eta_i g_{jk} + \eta_j g_{ki}
+\eta_k g_{ij} 
\end{align}
and $\nabla_iu_j=\nabla_ju_i$. Theorem \ref{thrm_2.12} states that $u$ is torse-forming and the perfect fluid spacetime is a GRW spacetime. Then $u^m C_{jklm}=0$ and 
$\nabla^m C_{jklm}=0$.\\
A contraction of \eqref{eq_RicciK} with $g^{jk}$ gives $\nabla_i R + 2\nabla_m R^m{}_i = (n+2)\eta_i$. However, it is $2\nabla_m R^m{}_i = \nabla_i R$. Then $\eta_k = \frac{2}{n+2} \nabla_k R$, where $R=nA-B$ is the curvature scalar.
\begin{prop}
For a perfect fluid spacetime with conformal Killing Ricci tensor
the equation of state is 
\begin{align}\label{eq_eqnstate}
p=-\mu \frac{n+1}{n-1} + \text{const.} 
\end{align}
\begin{proof}
Being $u$ an eigenvector of the Ricci tensor with eigenvalue $A-B$, by Lemma \ref{prop_2.10} it is $\eta_k=\nabla_k (A-B)$, then:
\begin{align}\label{eq_5.13}
\nabla_j[(n-2)A+nB] =0 
\end{align}
Since $B=8\pi (p+\mu)$ and $A=\frac{1}{2-n}8\pi (p-\mu )$, the equation of state is obtained.
\end{proof}
\end{prop}
\begin{remark} 
The condition $\nabla^m C_{jklm}=0$ applied to a perfect fluid gives the torse-forming vector
\eqref{eqnablau}, with $\gamma = (n-2) A + B$. The torse-forming property is the same as \eqref{eq_2.17}. This is ensured by 
$\nabla_m \gamma = (n-1)\nabla_m B$, because of $\nabla_m [(n- 2)A +nB ]=0$. The fact that a Ricci tensor may be conformal Killing is thus compatible with the conditions $\nabla^m C_{jklm}=0$, $u^m C_{jklm}=0$,
valid for perfect fluid GRW spacetimes.
\end{remark}

The equation of state \eqref{eq_eqnstate} with const.$=0$ violates the weak energy condition $|p/\mu|\le 1$. In $n=4$ it is 
 $p=-\frac{5}{3}\mu$. Matter with $p/\mu\le -1$ is named ``phantom energy''. It has positive energy density but negative pressure, such that $p+\mu<0$. The physical consequences are explored in \cite{Cal1,Cal2}.

%%%%%%%%%%%%%%%%%%%%%%%%%%%%%%%%%%%%%%%%%%%
%
\section{\bf Robertson-Walker spacetimes}
Robertson-Walker spacetimes are a subclass of GRW spacetimes and are conformally flat,  
$C_{jklm}=0$. Actually this property singles out RW spacetimes within GRW spacetimes:
\begin{thrm}[Brozos-V\'azquez et al, 2005, \cite{[BV]} theorem 1]\label{thrm_3.4}
A GRW spacetime $M$ is conformally flat if and only if $M^*$ is a space of constant curvature
(i.e. $M$ is a RW spacetime).
\end{thrm}
With $C_{jklm}=0$ in \eqref{eq_3.10}, the RW spacetime is a perfect fluid spacetime.
The Riemann tensor is then fully determined by Chen's vector, the metric tensor and the curvature scalar \cite{ManMol16a}:
\begin{align} 
&R_{jklm} =  \frac{2\xi-R}{(n-1)(n-2)} (g_{kl}g_{jm}- g_{km}g_{jl} ) \label{Riemann}\\
&+\frac{ R-n\xi }{(n-1)(n-2) } \left [ g_{jm} \frac{X_kX_l}{X^2} -g_{km} \frac{X_jX_l}{X^2} +
g_{kl}\frac{X_jX_m}{X^2}  - g_{jl} \frac{X_k X_m}{X^2} \right ] \nonumber  \end{align}
This form defines manifolds of quasi-constant curvature \cite{CheYan72}. 

\begin{remark}
The dimension $n=4$ is special, as the condition $X_m C_{jkl}{}^m=0$ is equivalent to $X_i C_{jklm}+X_j C_{kilm}+ X_k C_{ijlm}=0$ (see \cite{LovRun} p.128) and
the contraction with $X^i$ gives $C_{jklm}=0$. 
By the equivalence \eqref{equivalence}, in a 4-dimensional GRW spacetime the condition $\nabla_m C_{jkl}{}^m=0$ is equivalent to $C_{jklm}=0$. 
\end{remark}
\noindent
In agreement with \cite{GutOle09}, we state: 
\begin{thrm}
A four-dimensional GRW spacetime is a perfect fluid if and only if it is a RW spacetime. 
\end{thrm}
It is well known that a RW spacetime is a perfect fluid \cite{ONeill}. 
We recall Shepley and Taub's theorem for a 4-dimensional perfect fluid spacetime to be a RW spacetime:
\begin{thrm}[Shepley \& Taub, 1967, \cite{SheTau67}] \label{4.1}
A 4-dimensional perfect fluid spacetime with $\nabla_m C_{jkl}{}^m=0$ and subject to an
equation of state $p=p(\mu)$ is conformally flat, and the metric is RW; the flow is irrotational, shear-free and geodesic.
\end{thrm}
Other characterizations are the following:
\begin{thrm}[Coley, 1991, \cite{Coley91}]\label{thrm_4.3}
Any perfect fluid solution of Einstein's equations satisfying a barotropic equation of state $p=p(\mu)$,
$\mu+p\neq 0$, admitting a proper conformal Killing vector parallel to the velocity 4-vector $u_j$, 
is locally a Friedmann-Robertson-Walker model.
\end{thrm}
\begin{thrm}[Sharma, 1993, \cite{Sha93} p.3584] \label{thrm_4.4}
If a perfect fluid spacetime with divergence-free Weyl tensor admits a proper conformal
symmetry then it is conformally flat.
\end{thrm}
\begin{thrm}[Gutierrez \& Olea, 2009, \cite{GutOle09}, theorem 4.1] \label{thrm_4.5}
Let $M$ be a 4-dimensional non compact spacetime with a barotropic perfect fluid such that $u_j$ is geodesic,
$\partial p/\partial \mu \neq 0$,  $\partial^2 p/\partial \mu^2\neq 0$, $\mu+p\neq 0$, and $\mu>0$ is not constant. If the
equation of state 
\begin{align*}
\left[ \frac{\mu'}{\mu+p}\right ]' = \frac{1}{2}(3p+\mu) +\frac{1}{3}\left[ \frac{\mu'}{\mu+p}\right ]^2 
\end{align*}
holds, then either $M$ is incomplete or is a global RW spacetime.
\end{thrm}
The restriction $\nabla_m R_{jkl}{}^m=0$ is interesting.
In their review \cite{GuiNol98} of Yang's gravitational theory,  Guilfoyle and Nolan named ``Yang Pure Space" 
a 4-dimensional Lorentzian manifold ($M,g$) whose metric tensor solves Yang's equations:
\begin{align}
\nabla_k R_{jl}- \nabla_j R_{kl} =0 \label{eq_4.8}
\end{align}
In any dimension they are equivalent to $\nabla_m C_{jkl}{}^m =0$ and $R$ constant. If Einstein's 
equations \eqref{eq_Einstein} are considered, Yang's equations imply
$\nabla_k T_{jl} -\nabla_j T_{kl}=0$ and $\nabla_k T^m{}_m=0$. Therefore, in a perfect-fluid Yang Pure Space the 
equation of state is  $p=\tfrac{1}{3}\mu + \text{constant}$.
The following theorem was stated:
\begin{thrm}[Guilfoyle \& Nolan, 1998, \cite{GuiNol98} theorem 4.1]\label{thrm_4.2}
A 4-dimensional perfect fluid spacetime $(M,g)$ with $\mu+ p \neq 0$ is a Yang Pure Space if and only if $(M,g)$ is
a RW spacetime with $p=\frac{1}{3}\mu + c$ for some constant $c$.
\end{thrm}
For $n\ge 4$ it can be generalized:
\begin{prop}\label{5.5}
An $n$-dimensional perfect-fluid Yang Pure space with $p+\mu\neq 0$ is a GRW spacetime with equation of
state $p=\frac{\mu}{n-1}+ c$, being c a constant. 
\end{prop}
Finally we mention that Sharma and Ghosh proved that, for a $n=4$ expanding perfect fluid spacetime,
the energy momentum tensor $T_{ij}$ is conformal Killing if and only if the spacetime is shear-free,
vorticity-free, and $\mu $ and $p$ satisfy certain differential conditions \cite{ShaGho10}.

The case $R_{ij}=-R u_iu_j$ was discussed in \cite{ManSuhDe16}. Einstein's equations give 
$T_{ij} = \frac{R}{8\pi}(u_i u_j + g_{ij})$, i.e. the stress-energy tensor of a perfect fluid 
with $p=\mu$ ({\em stiff matter}, see \cite{Stephani} p.66). The equation of state of stiff matter was introduced by Zel'dovich \cite{Zel62} to describe a cold gas of baryons, and used in his cosmological model, \cite{Zel72}. 
The stiff matter era preceded the radiation era ($p=\mu/3$), the dust matter era ($p=0$), and the dark matter era ($p=-\mu$) \cite{Chavanis15b}. It also occurs in cosmological models where dark 
matter is a relativistic self-gravitating Bose-Einstein condensate \cite{Chavanis15a}.

A perfect fluid stress-energy tensor also arises in a spacetime with a real scalar field
with timelike gradient, $|\nabla \psi|^2= - g^{ij}(\nabla_i \psi)(\nabla_j \psi)$, 
(see \cite{HawEll,Wald}):
\begin{align*}
T_{ij}= (\nabla_i \psi)(\nabla_j \psi ) -\tfrac{1}{2}g_{ij} [V(\psi) -|\nabla \psi |^2 ],
\end{align*}
where $V(\psi)$ is a self-interaction potential that may have the simplest form 
$V(\psi) = \frac{1}{2}m^2 \psi^2$ ($m$ is the particle mass). 
A stiff matter model is recovered for a massless field (see \cite{Stephani} p.63). Setting 
$u_k = \nabla_k\psi / |\nabla\psi |$, we obtain  $T_{kl} = |\nabla\psi |^2 (u_ku_l +\tfrac{1}{2} g_{kl})$, with $u^2=-1$ and $p=\mu=\frac{1}{2}|\nabla \psi |^2$.
%
%%%%%%%%%%%%%%%%%%%%%%%%%%%%%%%%%%%%%%%%%%%%%%%%%%%%%%%%
%%%%%%%%%%%%%%%%%%%%%%%%%%%%%%%%%%
%
\section{\bf Other curvature conditions}
The following example illustrates several properties presented in this review. Mileva Prvanovi\'c  \cite{Prvan99} 
introduced the differential structure, named extended recurrence:
\begin{align}
\nabla_i R_{jklm} = & A_i R_{jklm}+ (\beta-\psi)A_i G_{jklm} \label{eq_prvd}\\
& +\frac{\beta}{2}[ A_j G_{iklm} + A_k G_{jilm} +A_l G_{jkim} +A_m G_{jkli}] \nonumber
\end{align}
where $A_i$ is a closed covector, $\psi $ and $\beta $ are scalar functions with $\nabla_i\psi =
A_i\beta$, and $G_{jklm} = g_{mj}g_{kl} - g_{km}g_{jl}$. Instead of the original Riemannian setup, we reconsider it as follows. 
\begin{prop}[\cite{ManMol16b}] An extended recurrent Lorentzian manifold with time-like vector $A_i$ is a conformally flat
GRW spacetime, i.e. a RW spacetime.
\begin{proof}
Contractions with the metric tensor of \eqref{eq_prvd} yield the expressions for $\nabla_i R_{jk}$ and $\nabla_i R$.
One finds out that the Weyl tensor is recurrent: $\nabla_i C_{jklm} = A_i C_{jklm}$.\\
The second Bianchi identity is readily written for \eqref{eq_prvd}. Being zero the l.h.s. a contraction of the r.h.s. with the metric tensor gives
\begin{align}
R_{jklm}A^m = [R_{jl}+\psi (n-2)g_{jl}] A_k - [R_{kl}+\psi (n-2)g_{kl}] A_j 
\end{align}
A further contraction shows that $A$ is eigenvector of the Ricci tensor:
$R_{jm}A^m = \frac{1}{2} [R+\psi (n-1)(n-2)] A_j $. A direct evaluation gives:
\begin{align*}
C_{jklm}A^m = \frac{n-3}{n-2} \left[ R_{jl}A_k -R_{kl}A_j + \left (\frac{R}{2(n-1)}+\psi \frac{n-2}{2} \right )
(A_kg_{jl}-A_j g_{kl}) \right ]
\end{align*}
Lovelock's identity (\cite{LovRun} p.289) is
\begin{align*} 
\nabla_i \nabla_m R_{jkl}{}^m + \nabla_j \nabla_m R_{kil}{}^m +\nabla_k \nabla_m R_{ijl}{}^m = -R_{im}R_{jkl}{}^m - R_{jm}R_{kil}{}^m - R_{km} R_{ijl}{}^m
\end{align*}
The evaluation of the l.h.s. gives zero, then
$R_{im} R_{jkl}{}^m +R_{jm} R_{kil}{}^m +R_{km} R_{ijl}{}^m = 0$. The covariant
divergence $\nabla^i$ of it, after lengthy calculations, gives:
$$ A_j \left[ R_{kl}- \frac{R}{2(n-1)}g_{kl} +\frac{n-2}{2}\psi g_{kl}\right ] =
A_k \left[ R_{jl}- \frac{R}{2(n-1)}g_{jl} +\frac{n-2}{2}\psi g_{jl}\right ] . $$
The result has nice implications:  $A_m C_{jkl}{}^m =0$ (then $\nabla_m C_{jkl}{}^m=0$), the contraction with $A^j$ gives: 
\begin{align}
R_{kl} =  \left[ \frac{R}{2(n-1)}-\frac{n-2}{2}\psi \right] g_{kl} + \left [\frac{n-2}{2(n-1)}R+\frac{n(n-2)}{2}\psi\right ] 
\frac{A_kA_l}{A^2} \label{Ricci_Mileva}
\end{align}
i.e. the space is quasi-Einstein.\\
The covariant derivative of the Ricci tensor shows that $A$ is a concircular vector (in the sense of Yano):
$\nabla_i A_j = f g_{ij}+\omega_i A_j$ with closed $\omega $. Then the metric has the warped form \eqref{eq_warped_Yano} with $F$ only depending on $t$, i.e. the Lorentzian manifold is a GRW spacetime. \\
A further step can be done. Since $\nabla_m C_{jkl}{}^m=0$, the second Bianchi 
identity for the Weyl tensor is exact:  $\nabla_i C_{jklm}+\nabla_j C_{kilm}+\nabla_k C_{ijlm}=0$.
Because of recurrence it becomes $A_iC_{jklm}+A_jC_{kilm}+A_kC_{ijlm}=0$. If $A^2\neq 0$, a contraction gives $C_{jklm}=0$ i.e.
the GRW spacetime is indeed RW.  
\end{proof}
\end{prop}
For a timelike vector $A_k$, on defining $u_k = A_k/\sqrt{-A^2} $, the Ricci tensor \eqref{Ricci_Mileva} is a perfect fluid.
Some physical consequences are outlined. Einstein's field equations give the stress-energy
tensor $T_{kl} =(p+\mu) u_k u_l +pg_{kl}$ with 
$$ p = -\frac{n-2}{16\pi (n-1)}[R+\psi (n-1)], \quad \mu =  -\frac{1}{16\pi} \psi (n-1)(n-2). $$
Thus the (non constant) function $\psi $ controls the energy density $\mu $ of the fluid, and must
be negative. The equation of state is:
$$  p=\frac{\mu}{n-1} - \frac{n-2}{n-1}\frac{R}{16\pi} $$
If $R=$ const.  the model is a perfect fluid Yang pure space (see Prop.\ref{5.5}).
In four dimensions with $R=0$, it becomes $p=\mu/3$, a model for incoherent radiation
(a superposition of waves of a massless field with random propagation directions, \cite{Stephani}).\\

Einstein GRW spacetimes are defined by $R_{ij}= \frac{R}{n}g_{ij}$.
From the general expression \eqref{eq_3.1} of the Ricci tensor  it easily follows that: 
\begin{prop}[S\'anchez, 1999, \cite{Sanchez99}]\label{thrm_3.1}
A GRW spacetime is Einstein if and only if the fibers are Einstein,
i.e. $R^*_{\mu\nu} = \frac{R^*}{n-1}g^*_{\mu\nu}$, and the warping function satisfies the differential equations
\begin{align}\label{eq_einsteinwarping}
\frac{R}{n(n-1)}=\frac{f^{\prime\prime}}{f}, \quad  \frac{R}{n(n-1)}f^2 = \frac{R^*}{(n-1)(n-2)} + (f')^2 .
\end{align}
\end{prop}
\noindent
The solutions are displayed in table 1 of \cite{Sanchez99} (see also \cite{[Ars14],AliRomSan97}).\\

We report some curvature properties, mainly investigated by R. Deszcz and collaborators. They are here adapted to GRW spacetimes.\\
The Riemann, the Ricci, the Weyl and the metric tensors on $M$
are denoted by ${\sf Riem,\, Ric,\, C}$ and ${\sf g}$; a star denotes tensors of the fiber $(M^*,g^*)$.

If ${\sf T}$ is a $(0,k)$ tensor and ${\sf A}$ is a $(0,2)$ symmetric tensor, the following $(0,k+2)$ tensors are
introduced: 
%Tachibana, Shun-ichi. Analytic tensor and its generalization. Tohoku Math. J. (2) 12 (1960), no. 2, 208--221.
%
\begin{align}
&({\sf Riem\cdot T})_{abc\dots lm} = \,- R_{lma}{}^pT_{pbc\dots } -
R_{lmb}{}^p T_{apc\dots } - R_{lmc}{}^pT_{abp\dots } - \dots \label{eq_3.21}  \\ 
&{\sf Q(A, T)}_{abc\dots lm} = \,
 A_{al}T_{mbc\dots } +A_{bl}T_{amc\dots } +A_{cl}T_{abm\dots } +\dots \label{eq_tachibana} \\ 
&\qquad\qquad\qquad\quad  - A_{am} T_{lbc\dots } - A_{bm}T_{alc\dots } -A_{cm} T_{abl\dots } -\dots \nonumber 
\end{align}
 ${\sf Q( A ,T)}$ is named Tachibana tensor.
In the product \eqref{eq_3.21} the Riemann tensor can be replaced by other curvature tensors. Note that in the present case:
$$({\sf Riem \cdot  T})_{lmab\dots } = -[\nabla_l, \nabla_m] T_{ab\dots}$$ 
The condition ${\sf Q(g,Riem)}=0$ is necessary and sufficient for a space to be of constant curvature, %Eisenhart
while ${\sf Riem \cdot Riem}=0$ defines semisymmetric manifolds. A generalization of semisymmetry is the following one, 
by Deszcz:
\begin{definition}[Deszcz \& Grycak, 1987]
A pseudo-Riemannian manifold is ``pseudosymmetric'' if, at every point, 
${\sf Riem\cdot Riem}$ and ${\sf Q (g,Riem )}$ are linearly dependent, 
i.e. there is a scalar function $\psi $ such that 
 \begin{align}
 {\sf Riem\cdot Riem} = \psi\, {\sf Q( g,Riem )}
 \end{align}
\end{definition} 
For example, the Einstein, RW ($\psi =1$), Schwarzschild, Kottler,
and Reissner-Nord\-str\"om spacetimes are pseudosymmetric.
\begin{thrm}[Deszcz, 1990,  \cite{Deszcz90}, theorems 3.4, 3.5]\label{thrm_3.13}
On a conformally flat manifold, ${\sf C}=0$, of dimension $n\ge 4$:\\
1) ${\sf Riem\cdot Riem = Q( Ric, Riem)} \Rightarrow $ $M$ is pseudosymmetric.\\
2) ${\sf Riem\cdot Riem =Q( Ric, Riem)}\Leftrightarrow$  $M$ is quasi-Einstein.
\end{thrm}
\begin{prop}[Defever \& al., 2000, \cite{DefDesHotKucSen00}, proposition 4.2]\label{thrm_3.14}
Let the warping function be $f^2(t)=a \exp (bt)$ with $a>0$ and $b\neq 0$, and $(M^*,g^*)$
such that: ${\sf C^*}=0$, $R^*=0$ and rank$({\sf Ric}^*)=1$. Then the following relations are satisfied:
\begin{align*}
& {\sf Riem\cdot Riem} = \frac{R}{n(n-1)} {\sf Q(g,Riem)}\\
& {\sf Riem \cdot Riem} -{\sf Q(Ric,Riem)} = - \frac{n-2}{n(n-1)} R\, {\sf Q( g,C)}\\
& {\sf Riem\cdot C} = \frac{1}{n-1} {\sf Q( Ric,C)}.
\end{align*}
\end{prop}
\begin{thrm}[Deszcz \& Kucharski, 1999, \cite{Desz_Kuch99}] 
In a $n=4$ GRW spacetime, if ${\sf Riem \cdot C}$ and ${\sf Q( Ric, C)}$ are
linearly dependent, then at least one of the two tensors ${\sf Riem\cdot C}$ or 
${\sf Q(Ric,C})$ must vanish.
\end{thrm}
In ref.\cite{[Ars14]}, Arslan et al. consider non-Einstein and non-conformally flat pseudo-Riemannian manifolds that, on the set 
$U = \{ x\in M : Q(S,R) \neq 0\}$ satisfy the condition
\begin{align}\label{eq_3.26}
{\sf Riem \cdot C - C\cdot Riem} = \,L \, {\sf Q(Ric ,Riem )}
\end{align}
where $L$ is some scalar function on $U$. They showed:
\begin{thrm}[Arslan \& al., 2014, \cite{[Ars14]}, theorem 4.1]\label{thrm_3.16}
Let $(M,g)$, $n\ge 4$, be the a GRW spacetime with non-Einstein fiber $(M^*,g^*)$. If \eqref{eq_3.26} holds, then  
$L=1/(n-2)$, the warping function is $f^2(t)=(at+b)^2$, $a,b\in\mathbb R$, 
\begin{align}\label{eq_3.28}
R^*_\alpha{}^\mu R^*_{\mu\beta\gamma\delta} = \frac{R^*}{n-1}
(R^*_{\alpha\beta\gamma\delta} + a^2 G^*_{\alpha\beta\gamma\delta}) 
- a^2 (g^*_{\beta\gamma}R^*_{\alpha\delta} - g^*_{\beta\delta} R^*_{\alpha\gamma}  )
\end{align}
\end{thrm}
The converse theorem has been proven: if $M^*$ is non-Einstein, if \eqref{eq_3.28} holds, and if
$f^2=(at+b)^2$, then the warped space $M=-1\times f^2 M^*$ fulfills \eqref{eq_3.26} with $L=1/(n-1)$. 

\begin{thrm}[Arslan \& al., 2014, \cite{[Ars14]}, theorem 3.1] \label{thrm_3.15}
On any Einstein manifold $n\ge 4$, \eqref{eq_3.26} holds with $L=1/(n-1)$.
\end{thrm}
If $(M,g)$ is Einstein and GRW,  then $f(t)= at+b$
(an admissible solution of \eqref{eq_einsteinwarping}).

%%%%%%%%%%%%%%%%%%%%%%%%%%%%%%%%%%

\section{\bf Remarks on imperfect fluid GRW spacetimes}
A large part of this survey dealt with perfect fluid GRW spacetimes, characterized by the condition $\nabla^m C_{jklm}=0$.
While RW spacetimes can only be perfect fluid (the Weyl tensor is zero), the general form \eqref{eq_3.10} of the Ricci tensor allows for GRW spacetimes solving the Einstein's field equations with a stress-energy tensor describing an imperfect fluid. Here we discuss the form \cite{Maartens}
\begin{align}
T_{ij} = (p+\mu) v_iv_j + p g_{ij} + P_{ij} \label{eq_TP}
\end{align}
where $v_i$ is the velocity of the fluid (a timelike unit vector field) and $P_{ij}$ is the anisotropic stress tensor: symmetric, traceless
and such that $P_{ij}v^j=0$.\\
We omit the possibility of heat transfer, that  adds a term $q_iv_j + q_j v_i$ to the tensor $T_{ij}$, with $q^iv_i=0$.
This term breaks the property that $v^j$ is an eigenvector of $T_{ij}$.
%On physical grounds $p$ and $\mu$ are related by an equation of state $p=p(\mu)$.
%The equations of motion $\nabla^k T_{kj}=0$ are (see \cite{Wald}):
%\begin{gather} \label{eq_4.3}
%u^k\nabla_k\mu + (\mu+p)\nabla^ku_k =0\\
%(\mu+p)u^k\nabla_k u_j +\nabla_j p + u_ju^k\nabla_k p=0
%\end{gather}
\begin{prop}
In a GRW spacetime the velocity field of a perfect fluid or an imperfect fluid described by the stress-energy tensor \eqref{eq_TP}, is torse-forming and proportional to Chen's vector.
\begin{proof}
Being an eigenvector of $T_{ij}$, the velocity $v$ is also an eigenvector of the Ricci tensor, by the Einstein's equations. In the warped frame of a GRW spacetime, the Ricci tensor has a block structure $R_{00}$ and $R_{\mu\nu}$, with
zero components $R_{0\mu}$, $R_{\mu 0}$. The tensor admits just one timelike eigenvector, the other $n-1$ being spacelike. Since this
counting is independent of the coordinate frame, the timelike eigenvector necessarily coincides with the torse-forming eigenvector  obtained by rescaling Chen's vector (theorem \ref{thrm_2.8}) $v=u$. 
\end{proof}
\end{prop}  
The property that the velocity is torse-forming implies that the acceleration $\dot u_i =u^k\nabla_k u_i$, 
the rotation tensor $\omega_{ij}=\frac{1}{2}h_i^r h_j^s (\nabla_s u_r-\nabla_r u_s)$  and the shear tensor
$\sigma_{ij} = \frac{1}{2}h_i^r h_j^s (\nabla_s u_r+\nabla_r u_s)- \frac{1}{n-1}h_{ij} \nabla_k u^k $ vanish. Since in the linear 
regime the standard constitutive relation for the anisotropic stress tensor is $P_{ij}=-\eta \sigma_{ij}$ \cite{Maartens}, it follows that
$P_{ij}=0$ in GRW spacetimes. \\
%The equations of motion $\nabla_kT^k{}_j=0$ simplify: 
%\begin{align}
%u_j [(n-1)\varphi (p+\mu) + \dot p + \dot \mu ]  + \nabla_j p + \nabla_k P^k{}_j =0, \label{nablaT}
%\end{align}
%with the definition $\dot x = u^k\nabla_k x$. \\
%The projection on $u^j$ gives: 
%$-(n-1)\varphi (p+\mu) - \dot \mu  + u^j\nabla_k P^k{}_j =0 $. However: $u^j\nabla_k P^k{}_j = \nabla_k( P^k{}_j u^j) - P^k{}_j
%\varphi (u_ku^j + \delta_k{}^j)=0$. We obtain the equation of energy conservation, 
%\begin{align}
%\dot \mu = -\varphi (n-1)(p+\mu)
%\end{align}
%equivalent to the formula $\nabla_k (u^k\mu) = - p \varphi (n-1)$. Eq. \eqref{nablaT} becomes:
%$ h_j{}^k\nabla_k p + \nabla_k  P^k{}_j  = 0 $.\\
If heat currents are allowed for, the velocity field no longer coincides with the torse-forming unit vector field (the latter, in GRW spacetimes is always an eigenvector of the Ricci tensor and an eigenvector of the stress-energy tensor).

%
%%%%%%%%%%%%%%%%%%%%%%%%%%%%%%%%%%%%%%%%%%%%
%

\end{document}